\documentclass[conference]{IEEEtran}
\IEEEoverridecommandlockouts
\usepackage{cite}
\usepackage{amsmath,amssymb,amsfonts}
\usepackage{graphicx}
\usepackage{textcomp}
\usepackage{xcolor}

\usepackage{amsmath, amssymb, amsthm,algorithm}
\usepackage[latin1]{inputenc}
\usepackage{xcolor}
\usepackage{latexsym}
\usepackage{graphics,epsfig}
\usepackage{amsfonts}
\usepackage{booktabs}
\usepackage{color}
\usepackage{multirow}
\usepackage[justification=centering]{caption}
\usepackage{graphicx}
\usepackage{adjustbox}
\usepackage{kantlipsum}
\usepackage[caption=false,font=footnotesize]{subfig}
\usepackage{cases}
\usepackage{url}
\usepackage{tabu}
\usepackage{makecell}
\usepackage{soul}

\usepackage{array}
\usepackage{algpseudocode}
\usepackage[justification=centering]{caption}

\usepackage[caption=false,font=footnotesize]{subfig}
\usepackage[T1]{fontenc}
\usepackage{mwe}
\usepackage{subfig}
\usepackage{mathtools,lipsum,cuted}

\usepackage[english]{babel}
\usepackage{verbatim}

\usepackage[english]{babel}

\def\BibTeX{{\rm B\kern-.05em{\sc i\kern-.025em b}\kern-.08em
    T\kern-.1667em\lower.7ex\hbox{E}\kern-.125emX}}
\begin{document}

\title{Increasing Fault Tolerance and Throughput with Adaptive Control Plane in Smart Factories}
\author{\IEEEauthorblockN{Cao Vien Phung and Admela Jukan}
\IEEEauthorblockA{Technische Universit\"at Braunschweig, Germany\\
Email: \{c.phung, a.jukan\}@tu-bs.de}}
\maketitle

\begin{abstract}
Future smart factories are expected to deploy an emerging dynamic Virtual Reality  (VR) applications with high bandwidth wireless connections in the THz communication bands, where a factory worker can follow activities through 360\textdegree video streams with high quality resolution. THz communications, while promising as a high bandwidth wireless communication technology, are however known for low fault tolerance, and are sensible to external factors. Since THz channel states are in general hard to estimate, what is needed is a system that can adaptively react to transceiver configurations in terms of coding and modulation. To this end, we propose an adaptive control plane that can help us configure the THz communication system.  The control plane implements a workflow algorithm designed to adaptively choose between various coding and modulation schemes depending on THz channel states. The results show that an adaptive control plane can improve throughput and signal resolution quality, with theoretically zeroed bit error probability and a maximum achievable throughput in the scenarios analayzed. 
 \end{abstract}

\section{Introduction} \label{intro}
\begin{figure*}[ht]
  \centering
  \subfloat[]{\includegraphics[ width=4.3cm, height=4.6cm]{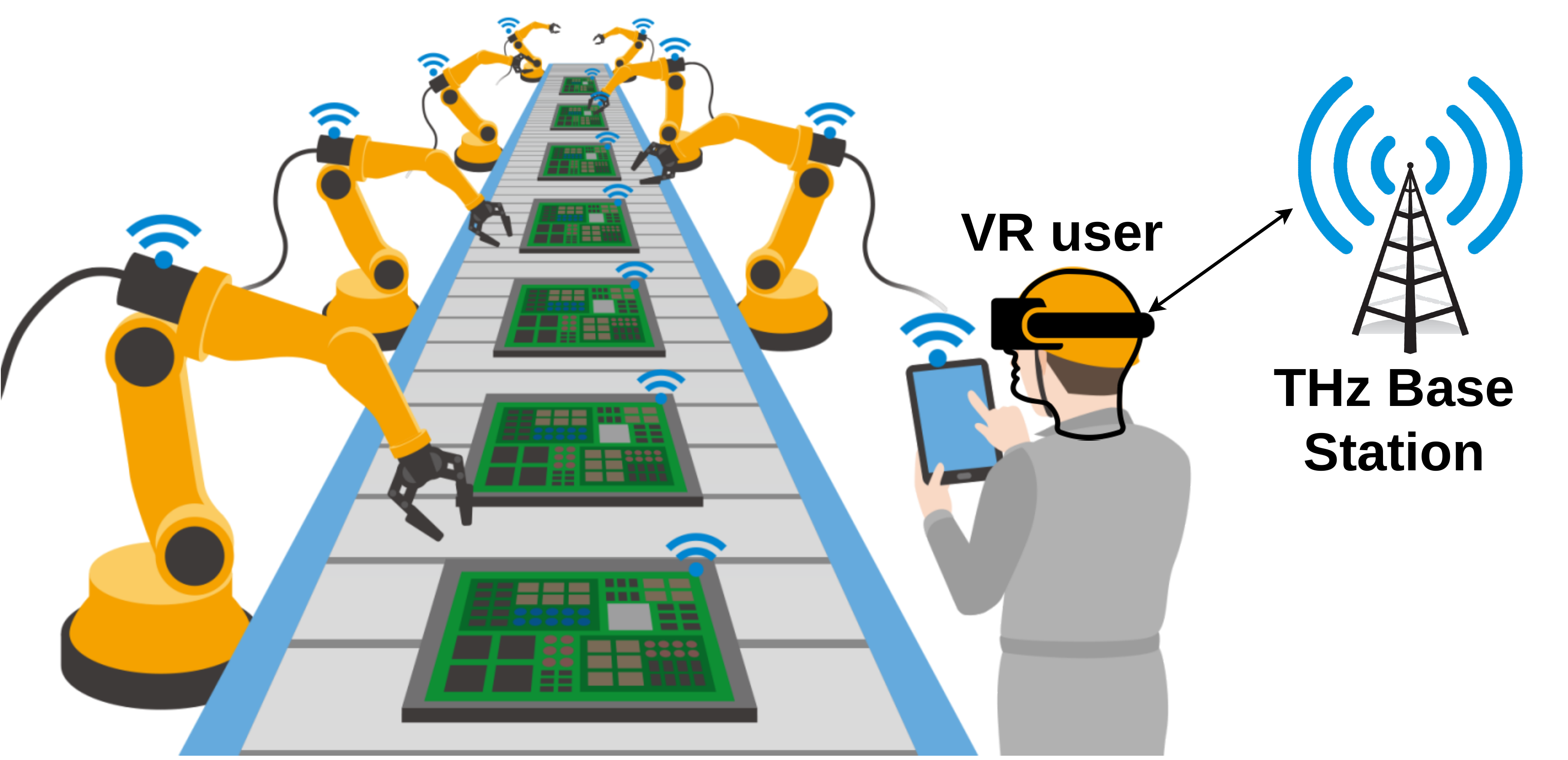}
  \label{fig:vr_scenario}}
  \subfloat[]{\includegraphics[ width=13.8cm, height=5.4cm]{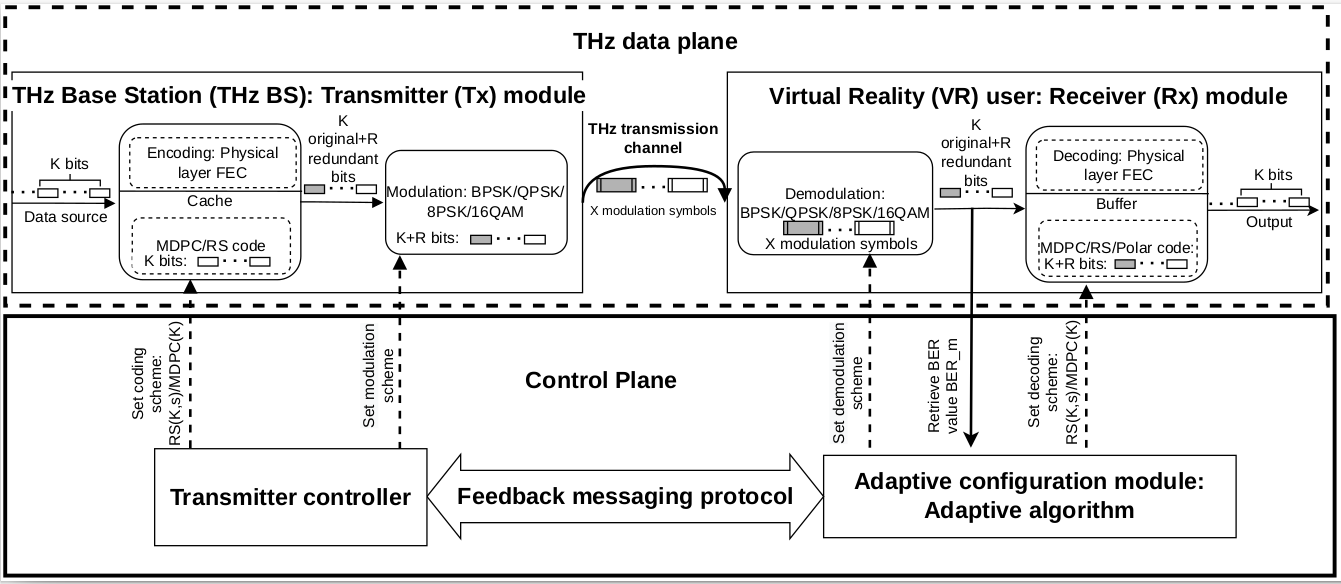}
  \label{fig:arch}}
  \caption{(a) An augmented reality smart factory application with wireless communications in THz band; (b) A THz control plane architecture with feedback loop workflow protocol.}
  \label{results}
  \vspace{-0.5cm}
  \end{figure*}

Future smart factory scenarios are envisioned with the downlink applications using high bandwidth wireless links able to provide high throughput and fault tolerance. A typical scenario, as illustrated in Fig. \ref{fig:vr_scenario}, includes several robots, a factory worker is using an augmented or virtual reality (VR) and high quality resolution device (as a receiver), and a Base Station (BS) (as a transmitter).  The high data rate of communications in Gb/s is today achievable with Terahertz (THz) frequency band ($0.3-10$ THz). A system that can combine THz communications, and virtual- and augmented reality will enable several advanced features in the manufacturing process, such as  the real-time placement of images onto the real working environment to see  the proper instructions on how to assemble a particular product or a component, hands-free \cite{boeing2018}. Such applications in general are expected to improve the manufacturing efficiency, decrease the training time of factory works, and enhance the safety.

To guarantee high throughput and fault tolerance, the system needs to be able to adaptively respond to changed transmission and applications conditions. First, this is necessary since VR users move, leading to the outdated channel states, making static system configuration obsolete.  Second, and more critically, THz band is sensitive to molecular absorption \cite{7444891}, leading to an easily deteriorated Signal-to-Noise Ratio (SNR), i.e., a high Bit Error Rate (BER). Thus, to provide the desired robustness, the adoption and development of the adequate coding schemes are subject of ongoing research \cite{2018_WEHN.SAHIN_NextGenerationChannelCoding}. In regard to coding, a few FEC codes with redundant data are under discussion for THz communication to protect original data against losses, such as Low Density Parity Check (LDPC), RS or Polar code \cite{wehn_norbert_2018_1346686, wehn_norbert_2019_3360520}. To increase the transmission performance, data can also be transferred with either a higher modulation formats, or over longer paths, which also is a major challenge in THz spectrum. Previous work analyzed time-varying channel modeling and tracking for THz indoor communications \cite{8292607}. Papers \cite{9013838} and \cite{DBLP:journals/twc/PengGK18} analyzed to this end different mobility scenarios in THz indoor communications.  

In this paper, we conceptualize the need to provide an adaptive system configuration in response to mobility and channel state changes, with the aim of improving throughput and fault tolerance. To this end we propose data and control plane architectures, akin to what is known in high speed networks, e.g., \cite{Chamania2020}. In the control plane, we design a workflow feedback loop algorithm which can adaptively update the channel states and generate adaptive coding and modulation THz system configurations. In the data plane, we consider four modulation schemes: BPSK, QPSK, 8PSK and 16QAM.  For coding, the system can choose between and Multidimensional Parity Check (MDPC) \cite{8867037} and Reed-Solomon (RS) codes \cite{Reed1999}. The reasons for the above selections is because these modulations and coding schemes are well known and can be analyzed, though our approach is applicable to other schemes as they emerge. The results show an adaptive control plane can choose the best coding and modulation schemes with the lowest transmission overhead, and the highest code rate and throughput, therefore improving the system performance with theoretically zeroed bit error probability and a maximum achievable throughput in the scenarios analayzed.

\vspace{-0.1cm}

\section{System Design}\label{sce}

\subsection{THz data plane}
The THz data plane includes the transmitter (Tx), also referred to as THz base station, and the receiver (Rx) module, here corresponding to the VR device. In our model we assume that Rx (VR) can move, whereas Tx (BS) is static.  At the Transmitter (Tx) (THz base station), the data source with a long bit stream is split into substreams. Each substream is referred to as a coding generation of $K$ bits, whereby the coding generation includes a set of bits coded and decoded together. Based on channel condition, $K$ can be adaptively chosen and all Ks are temporarily stored at the coding cache for the coding process. A coding generation of $K$ bits generates $R$ redundant bits from the coding process, where $R$ bits are used to support for correcting error bits from $K$ bits. We consider two coding schemes: RS code, and MDPC($n$D/$m$L) code with $n$ dimensions (D) and the same length (L) of $m$ bits for each dimension, whereby the original bits stay unchanged after coding, meaning that only redundant coded bits are generated by coding. The control plane can choose one of these two schemes adaptively, based on an algorithm.  

If MDPC($n$D/$m$L) code is chosen, the encoding process for $R=(m+1)^n - m^n$ parity bits \cite{8867037} generated from $K=m^n$ bits can be performed as follows: A special case of MDPC($n$D/$m$L) code with n = 2 has column and row parity bits in the last row and column of the two-dimensional matrix, respectively. The original data bits are put at the remaining columns and rows of the matrix. Data bits in the column of the matrix and in the row of the matrix are secured by one column parity bit in the last row and by one row parity bit in the last column, respectively. The parity check bit on check bits for securing column and row parity bits is at the bottom right corner of the matrix. The construction of MDPC($n$D/$m$L) code with $n \geq 3$ is a multidimensional hypercube combined from basic three-dimensional parity check cubes. A parity check cube of three dimensions is a set of the two-dimensional parity matrices layered into a third dimension and the parity bits across the layers of data bits belong to the last layer.

If RS code is chosen, the related coding mechanism as well known from \cite{Reed1999} can be performed as follows: The original data $M(X)$ with the size of $K$ bits is operated on the coding process with the modulo-$g(X)$ function to create the redundant data $CK(X)$, i.e., $CK(X)=X^{\frac{R}{s}}\cdot M(X) \; mod \; g(X)$, whereby $X^{\frac{R}{s}}$ represents the displacement shift, $s$ is the coding symbol size in bits, $R$ is the size of redundant coded information bits and $g(X)$ represents the generator. Therefore, the RS code word $C(X)=X^{\frac{R}{s}}\cdot M(X)+CK(X)$ includes the redundant data $CK(X)$ appended systematically onto the original data $M(X)$, i.e., $K$ original bits are increased to $K+R$ bits during the coding process. $R+K$ bits are divided into symbols of $s$ bits each and all coding operations are performed via finite field $\mathbb{F}_{2^s}$. The maximum codeword length of RS code is $\frac{K+R}{s}=2^s-1$. If $R$ is fixed and $\frac{K+R}{s}<2^s-1$, then $Z$ zero padding symbols can be added into $\frac{K}{s}$ original symbols by the encoder so that the codeword length $\frac{K+R}{s} + Z=2^s-1$ can be achieved, and $\frac{R}{s}$ redundant coded symbols are generated. These zero padding symbols will not be sent out of transmission channels, but the decoder needs to use them for the decoding process. After the coding process, $K+R$ data bits are modulated to generate $X$ modulation symbols. The control plane can choose one of the four modulation schemes: BPSK, QPSK, 8PSK and 16QAM, depending on the THz channel state. After the modulation process, $X$ modulation symbols are sent over the THz channel. 

At the Receiver (Rx) side (VR device), we collect $X$ symbols to demodulate.  $X$ modulation symbols are demodulated to re-create $K$ original data bits and $R$ redundant bits, for decoding. $K+R$ bits will be then delivered to the decoding block. As large amount of data arrives at the same time, a receiving buffer is necessary to temporally store all arriving data for the decoding process, whereby the decoded generations are serialized. If MDPC($n$D/$m$L) code is used, the iterative decoding method can be applied \cite{8867037} to  correct the data bits with the highest error probability in each decoding iteration. All parity bits and their related data bits are checked in each decoding iteration, and then we will count a failed dimension marker (FDM). The FDM of the arbitrary bit is set to $k$, if that bit discovers a parity check error in the $k$ dimension. If the FDM reaches the highest value less than $2$, then the decoding process stops.  Otherwise, all bits are inverted with the highest FDM value. In the next step, the iteration number limit is checked to decide whether the decoding process stops or all steps are repeated in the next iteration, based on  the
predefined maximum iteration. Finally, if $FDM_{max}=0$ after the decoding process,  the frame is considered
error-free. Otherwise, we face an uncorrectable error.

If RS code \cite{Reed1999} is chosen, the data of RS code is decoded as follows. The syndrome components are calculated from the received word at the first stage. The second stage is to calculate the error-locator word from the syndrome components. At the third stage, the error locations are calculated from the error-locator numbers which are from the error-locator word. The fourth stage is for calculating the error values from the syndrome components and the error-locator numbers. Finally, we calculate the decoded code word from the received word, the error locations, and the error values.

\begin{figure}[!t]
    \centering
    \includegraphics[scale=0.4]{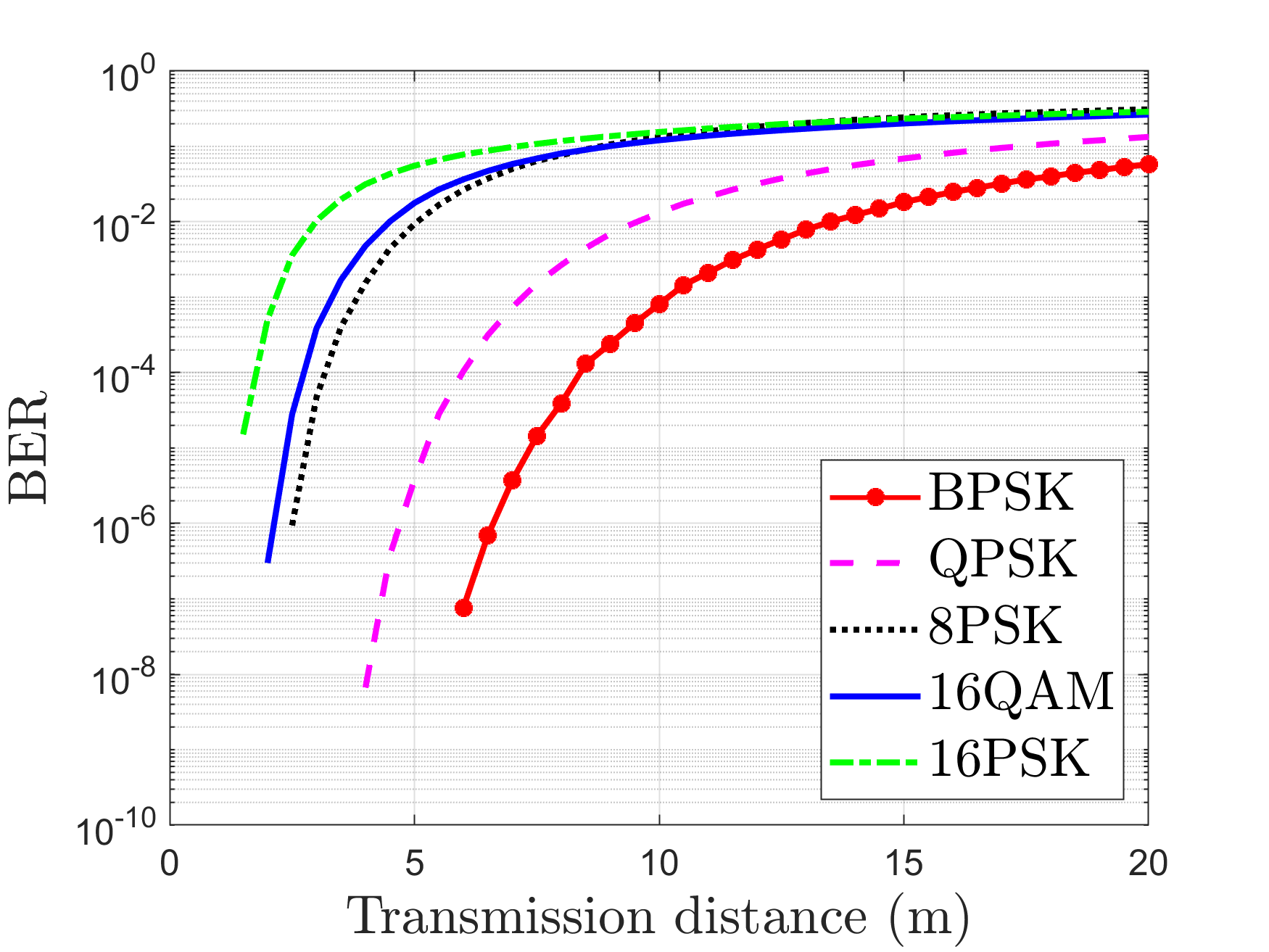}
    \caption{BW 8.64 GHz at center
frequency 300.24 GHz.}
    \label{BER_1ch_BW8_64GHz}
  \vspace{-0.7cm}
\end{figure}

\subsection{THz Control Plane}
The control plane module includes the two main components referred to as Transmitter Controller and Adaptive Configuration Module. In between, a messaging protocol (feedback messaging protocol) is used for communications over a wireless channel. The latter can be either in THz or out of THz band, and is not critical in the overall design due to its low bandwidth required.

Adaptive configuration module generates new system configurations when the channel state changes. This module collects the necessary measurements (e.g., BER) used for estimating the receiver performance as well as channel state information. Such measurements are not foreseen at the transmitter module, here THz BS, but can be easily added to the system. For instance, transmitter can measure the noise levels and calibrate the transmitting power as the noise increases (e.g., due to equipment aging or heat). The arrow labelled with the message "Retrieve BER value $BER_m$" illustrated that the measured BER value is periodically updated from the receiver module to the adaptive configuration module. The messaging arrows labelled as "Set demodulation scheme" and "Set decoding scheme: RS($K,s$)/MDPC($K$)" illustrate the actuation messaging to inform the data plane when the best demodulation and decoding schemes are chosen, respectively. This messages are sent out after the new system configurations are generated by the adaptive configuration module.

It is important to note that a buffer needs to be defined sufficiently large for the system to work and store the messages about the system states. Let us define a buffer $B$ with an enough size for $N$ messages. This buffer is used to save measured BER values $BER_m$ (before decoding) sent by the receiver. $BER_m$ after buffering is denoted to be $BER_f$. The $BER_fs$ stored in the buffer are used to anticipate the status of the receiver (e.g., moving or standing idly at a certain position), or the transmission distance. For example, changed $BER_fs$ mean that it is likely that the receiver is moving. Consider the BER values of bandwidth $8.64$ GHz at center frequency $300.24$ GHz over different transmission distances and modulation schemes conducted using the Simulator for Mobile Networks (SiMoNe) in Fig. \ref{BER_1ch_BW8_64GHz} from \cite{9391508}, e.g., if we have $BER_m$ or $BER_f$ is approximately  equal to $0.0579$, then the transmission distance is estimated to be $d=20$ m with using BPSK, whereby $BER_m$ and $BER_f$ can be deduced for the current transmission distance and previous transmission distance, respectively. At the beginning, we store the value of $0$ in the buffer. The values $BER_m$ are periodically updated from the receiver module at a constant time interval $t$. We assume that a small moving speed of a pedestrian is at $1$ m/s, i.e., a footstep of worker with $0.5$ m will take $0.5$ s. So, to estimate the moving of receiver, we recommend setting the constant time interval $t=0.5$ s for periodically updating $BER_m$.

In Fig. \ref{BER_1ch_BW8_64GHz}, on the other hand, the fluctuations of channel state are continuous when the receiver is moving to different transmission distances. Hence, the adaptive configuration module only generates the new configurations, when the receiver is predicted to stop at a certain position. This can be explained as follows: The channel state changes so fast, when the receiver is moving. So, the new configurations have not been able to actuate yet, and are already obsolete.  Also, too frequent channel state update can cause an extremely large overhead in each transmitted frame. At the beginning, let us assume that we set up the system using RS code $(224,240)$ and the main THz channel using 16QAM, and they are known by transmitter, receiver and adaptive configuration module. 

\subsection{Adaptive Algorithm}\label{algo}
To generate new states, an Algorithm \ref{algorithm1} is used. For line \ref{sel}, the Selection function includes $BER_m$ periodically updated by the receiver module. For line \ref{if1}-line \ref{endi}, we compare $BER_m$ with the other previous BER values $BER_f$s stored in the buffer. If there exists at least a difference of $BER_m$ and any $BER_f$ stored in the buffer $B$, where that difference is more than or equal to $ \varepsilon $, then it is likely that the receiver is moving, whereby $\varepsilon>0$. For example, based on BER values over different distances and modulation schemes in Fig. \ref{BER_1ch_BW8_64GHz}, if the difference between $BER_m$ and $BER_f$ is at least $ \varepsilon \approx  7.646\cdot 10^{-8}$ for BPSK, $\varepsilon \approx  6.649 \cdot 10^{-9}$ for QPSK, $\varepsilon \approx  9.375 \cdot 10^{-7}$ for 8PSK and $\varepsilon \approx  2.992 \cdot 10^{-7}$ for 16QAM for bandwidth $8.64$ GHz at center frequency $300.24$ GHz, then the receiver is predicted to be moving. When the receiver is moving, $BER_m$s will be continuously changed. Therefore, $BER_f$s previously stored in the buffer will be removed, and then  the new $BER_m$ is added into the buffer. The reason for the deletion is that when the receiver is moving, $BER_f$s become outdated. If $\left| BER_m -BER_f  \right| < \varepsilon$, $\forall BER_f \in B$ (line \ref{el}), then it is likely that the receiver stops at a certain position. However, no new configuration is generated because we are not totally sure that whether the receiver keeps walking or not, i.e., we need more $BER_m$ updates to give a decision of generating new configurations. If the buffer $B$ is not full (line \ref{fu}), then we add the new $BER_m$ into it (line \ref{ad}). We keep adding new $BER_m$s, which satisfy the condition of line \ref{el}, until the buffer is full (line \ref{ful}).

For line \ref{dis}-\ref{resu}, the adaptive configuration module starts generating new adaptive configurations because highly likely the receiver stops moving. Based on $BER_m$ received with the current modulation scheme used by the transmitter and the BER database over different transmission distances and modulation schemes stored in this module, we can approximately predict the current transmission distance between the transmitter and receiver (line \ref{dis}). Based on the current distance of THz transmission channel and the BER database over different transmission distances and modulation schemes stored in this module, we can approximately predict the bit error probability $p_e$, if the transmitter uses modulation scheme $e \in \{$ BPSK, QPSK, 8PSK,16QAM $\}$ at that transmission distance of main THz channel (line \ref{pe}). Next, we discuss line \ref{resu}. First of all, we consider the number of error bits $t_{MDPC}$ corrected by MDPC code and the number of error symbols $t_{RS}$ corrected by RS code: 
 \begin{equation}\label{tv}
\text{      } \left\{\begin{matrix}
t_{MDPC}=2^{n-1} - 1, & \\ 
t_{RS}=\frac{R}{2s}&,
\end{matrix}\right.
 \end{equation}
whereby $t_V$ ($V \in $ \{MDPC code, RS code\}) is assumed to be fixed. On the other hand, we consider two parameters: Code rate $R_F$ and coding system throughput $TH$ defined to be the number of useful data bits transferred by the THz transmission system to a certain destination per unit of time. These two ones with formulas are expressed by:
 \begin{equation}\label{pa-rf-th}
\text{      } \left\{\begin{matrix}
R_F=\frac{K}{K+R}, & \\ 
TH=R_F \cdot D \cdot (1-P_{re}^V), &
\end{matrix}\right.
 \end{equation}
where $D$ is the data transmission rate and $P_{re}^V$ is the residual bit error probability after decoding. With fixed value $t_{MDPC}$ (Eq. \eqref{tv}) and $p_e$ from four modulation schemes considered from line \ref{pe}, we can choose $4$ optimized values of $K=m^n$ original bits with $R=(m+1)^n-m^n$ redundant bits from $4$ modulation schemes for MDPC code, where $m,n \in \mathbb{N}$, so that they satisfy the eqution as:
\begin{equation}\label{parameters-mn}
(m+1)^n \cdot p_e\leq t_{MDPC}.
 \end{equation}
At the same time, similarly, with fixed value $t_{RS}$ (Eq. \eqref{tv}) and $p_e$ from four modulation schemes considered from line \ref{pe}, we can choose $4$ optimized values of $K$ original bits with $R$ redundant bits from $4$ modulation schemes for RS code, so that they satisfy the eqution as:
 \begin{equation}\label{parameters-RKforRS}
\text{      } \left\{\begin{matrix}
\frac{K+R}{s} \cdot  P_s   \leq t_{RS}\; \; ;\; \; K\geq 0 & \\ 
2^{s-1}\leq \frac{K}{s}+\frac{R}{s} \leq 2^s - 1 &,
\end{matrix}\right.
 \end{equation}
where $P_s=1-(1-p_e)^{s}$ is the symbol error probability on the main THz channel and the maximum codeword length of RS code is $\frac{K+R}{s}=2^s-1$. Note that the values of $K$, $R$ or $s$ chosen from Eq. \eqref{parameters-mn} for MDPC code and Eq. \eqref{parameters-RKforRS} for RS code over different modulation schemes, the total number of error bits/symbols occurred is lower than or equal to the ability of correcting bit/symbol errors by MDPC($n$D/$m$L) and RS code, when $K+R$ bits or $\frac{K+R}{s}$ symbols are sent over the main THz channel with the bit error probability $p_e$ or symbol error probability $P_s$, with the aim of optimizing the fault tolerance system. Theoretically, with this approach, the residual bit error probability $P_{re}^V$ after decoding is approximately equal to $0$. From eight optimized values of $K$ original bits/symbols with $R$ redundant bits/symbols of MDPC and RS code over four modulation schemes, we can choose the best coding and modulation scheme with the highest code rate $R_F$ and throughput $TH$, based on Eq.\eqref{pa-rf-th}. As the adaptive configurations are chosen, the THz data transmission can achieve high throughput and code rate with an optimized system fault tolerance. If $BER_m$ arrives when the buffer of adaptive configuration module is full and $\left| BER_m -BER_f  \right| < \varepsilon$, $\forall BER_f \in B$ (line \ref{ell}), then we only remove the oldest $BER_f$ stored in the buffer, and then add $BER_m$ into it. In this case, we do not generate any new system configurations because the receiver does not still change its position, i.e., the current channel state is similar to the previous one with the adaptive configurations already generated from line \ref{ful} to line \ref{resu}. Note that if MDPC code is chosen, this module generates the configuration: MDPC($K$), whereby $K$ is the size of original input data. Else if, RS code is chosen, this module generates the configuration: RS($K,s$), whereby $s$ is the size of coding symbol in bits.

\subsubsection{Analysis of algorithm complexity} \label{complexityvien}
 
In this subsection, we calulate the overhead for generating a new adaptive configuration, when a $BER_m$ is updated at the adaptive configuration module. As analyzed from line \ref{fu}-line \ref{resu} in Algorithm \ref{algorithm1}, this module only generates new configurations, if the new $BER_m$ arrives at the time when the buffer $B$ stores $N-1$ $BER_f$ messages. Assume that a computation for the process of generating adaptive configurations is counted as one unit and we ignore the simple processes of removing or storing from the bufffer. We need $N-1$ units to compare $BER_m$ and $BER_f$ at line \ref{if1}. At line \ref{dis}, we need one unit for predicting the transmission distance of VR user. At line \ref{pe}, we need $M=4$ units for predicting the bit error probability for four modulation schemes. Finally, for each coding scheme, we need $M \cdot 3=12$ units at line \ref{resu} because we have $M=4$ values of bit error probability $p_e$ applied for $3$ equations (Eqs. \eqref{pa-rf-th}, \eqref{parameters-mn}, \eqref{parameters-RKforRS}) to find the best coding and throughput with the highest code rate and throughput. As a result, in general, we need $N+M(1+3C)$ units for generating a new system configuration, where $M$ is the number of modulation schemes considered and $C $ is the number of coding schemes considered.

  \begin{algorithm}
\caption{Adaptive Workflow Configuration Algorithm}
\label{algorithm1}
\begin{algorithmic}[1]
\Function{selection}{$BER_m$} \label{sel}
\If{ $\exists$  $\left| BER_m -BER_f  \right| \geq \varepsilon$, $\forall BER_f \in B$ } \label{if1}
\State Remove all $BER_f$s, and buffering $BER_m$;\label{endi}
\Else \label{el}
\If{$B$ is not full} \label{fu}
\State Add $BER_m$ into the buffer; \label{ad}
\If{$B$ is full} \label{ful}
\State Based on $BER_m$ $\rightarrow$ transmission distance.\label{dis}
\State Based on transmission distance $\rightarrow$  bit error probabilities $p_e$, if using BPSK, QPSK, 8PSK or 16QAM.\label{pe}
\State Using $p_e$ from four modulation schemes, $t_V$ fixed and Eqs. \eqref{pa-rf-th}, \eqref{parameters-mn}, \eqref{parameters-RKforRS}  $\rightarrow$ chosen coding and modulation scheme with the highest $R_F$ and $TH$. \label{resu}
\EndIf
\Else \label{ell}
\State Remove the oldest $BER_f$, and buffer $BER_m$;\label{remd}
\EndIf
\EndIf
\EndFunction
\end{algorithmic}
\end{algorithm}

\begin{figure*}[ht]
  \centering
  
  \subfloat[Transmission distance.]{\includegraphics[ width=5.9cm, height=4.2cm]{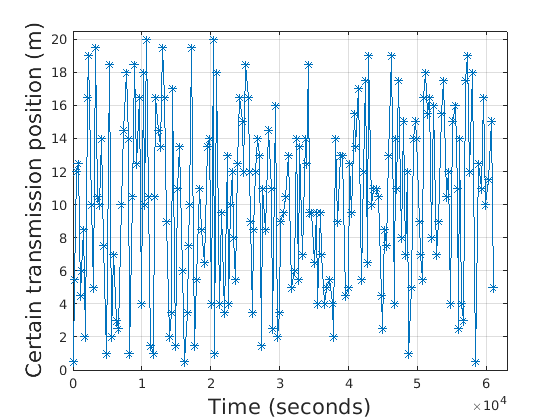}
  \label{distance}}
  \subfloat[Code rate.]{\includegraphics[ width=5.9cm, height=4.2cm]{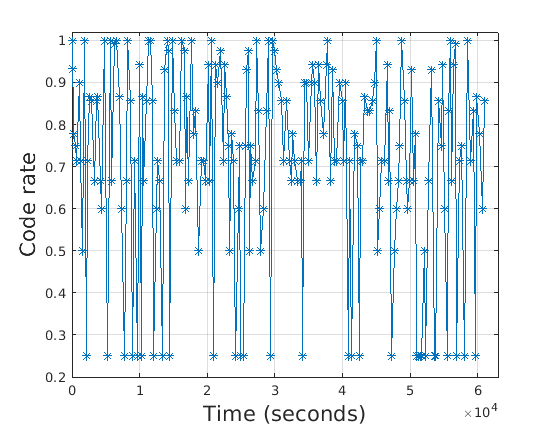}
  \label{coderate}}
  \subfloat[Transmission overhead.]{\includegraphics[ width=5.9cm, height=4.2cm]{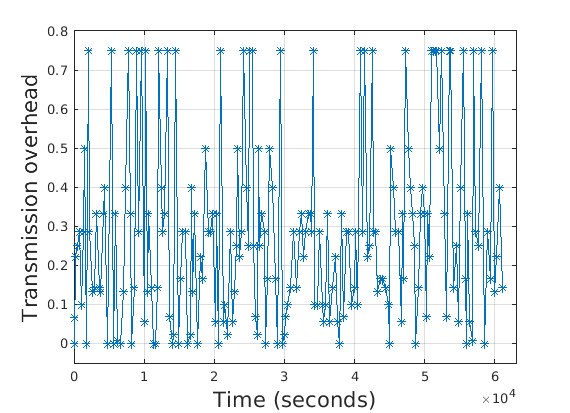}
  \label{transoverhead}}
  \caption{BER vs. distance without coding, transmission distance randomly chosen, code rate and transmission overhead.}
  \label{results}
  \vspace{-0.5cm}
  \end{figure*}
  
  \begin{figure*}[ht]
  \centering
  \subfloat[Chosen coding scheme.]{\includegraphics[ width=5.9cm, height=4.2cm]{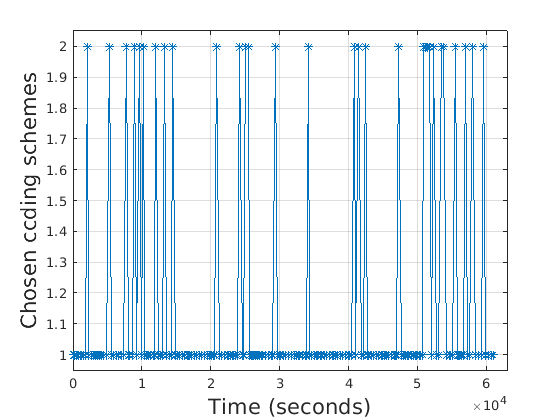}
  \label{codingschemechosen}}
  \subfloat[Chosen modulation scheme.]{\includegraphics[ width=5.9cm, height=4.2cm]{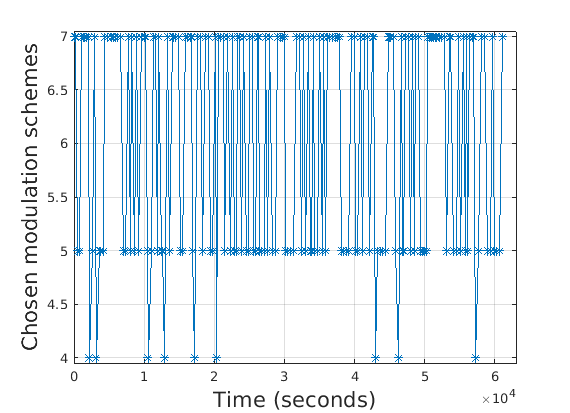}
  \label{modulationschemechosen}}
  \subfloat[Throughput.]{\includegraphics[ width=5.9cm, height=4.2cm]{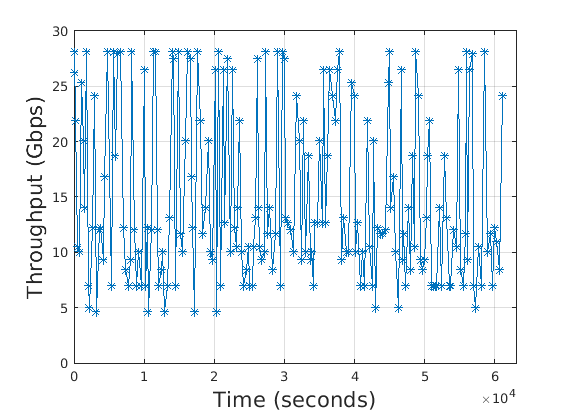}
  \label{throughputTH}}
  \caption{Theoretical results. For chosen coding scheme,  $1$ and $2$ denote chosen RS code and MDPC code, respectively. For chosen modulation scheme, $4,5,6$ and $7$ represent chosen BPSK, QPSK, 8PSK and 16QAM, respectively..}
  \label{results}
  \vspace{-0.5cm}
  \end{figure*}
 \section{Performance Evaluation}
In this section, we show the performance results in the smart factory applications shown, considering THz data and control planes proposed.  All data including over the THz channel, which has bandwidth $8.64$ GHz at center frequency $300.24$ GHz. For the BER values, we make use of the data shown in Fig. \eqref{BER_1ch_BW8_64GHz}. 
These BER values are used to estimate the transmission distance between VR user and THz BS and  bit error probability $p_e$ over different modulations under the Algorithm \ref{algorithm1}. To estimate the moving of VR user, we assume setting the constant time interval $t=0.5$ s for periodically updating $BER_m$ from VR user to the adaptive configuration module. Note that at the beginning, we store temporally the value $0$ in the buffer and use RS code $(224,240)$ with the modulation scheme of 16QAM. The adaptive configuration module only generates the new configurations, when VR user stops at a certain position. As discussed in Algorithm \ref{algorithm1}, based on the database of BER values from Fig. \ref{BER_1ch_BW8_64GHz}, we define $ \varepsilon \approx  7.646\cdot 10^{-8}$ for BPSK, $\varepsilon \approx  6.649 \cdot 10^{-9}$ for QPSK, $\varepsilon \approx  9.375 \cdot 10^{-7}$ for 8PSK and $\varepsilon \approx  2.992 \cdot 10^{-7}$ for 16QAM to predict whether the VR user is moving or stops at a certain place. The data transmission rate $D$ for BPSK, QPSK, 8PSK and 16QAM is $7.04$ Gbps, $14.08$ Gbps, $21.12$ Gbps and $28.16$ Gbps, respectively.  

\begin{table}[t!]
  \centering
  \caption{List of main values applied for evaluation.}
  \label{tab:table1}
  \begin{tabular}{ll}
    \toprule
   Values & Meaning\\
    \midrule
    $t_{MDPC}=1$ & Number of bits can be corrected by MDPC code.\\
$t_{RS}=1$ & Number of symbols can be corrected by RS code.\\
$N=4$ & Maximum number of messages stored in buffer $B$.\\
$t=0.5$ & Time for periodically updating $BER_m$ in second.\\
$ \varepsilon \approx  7.646\cdot 10^{-8}$ & Likely VR user is moving if using BPSK.\\
$\varepsilon \approx  6.649 \cdot 10^{-9}$ & Likely VR user is moving if using QPSK.\\
$\varepsilon \approx  9.375 \cdot 10^{-7}$ & Likely VR user is moving if using 8PSK.\\
$\varepsilon \approx  2.992 \cdot 10^{-7}$ & Likely VR user is moving if using 16QAM.\\
$D=7.04$ & Data transmission rate for BPSK in Gbps.\\
$D=14.08$ & Data transmission rate for QPSK in Gbps.\\
$D=21.12$ & Data transmission rate for 8PSK in Gbps.\\
$D=28.16$ & Data transmission rate for 16QAM in Gbps.\\
     \bottomrule
  \end{tabular}\vspace{-0.5cm}
\end{table}

Fig. \ref{distance} shows the different standing positions (transmission distances) of VR user around $60600$ s, where they are randomly chosen between $0.5$ m and $20.0$ m in steps of $0.5$ m from the transmitter. When VR user moves between two positions, the channel state changes. Assume that based on $BER_m$ periodically updated from the VR user (line \ref{dis} in Algorithm \ref{algorithm1}). Based on the channel states in Fig. \ref{distance}, our adaptive algorithm can choose the best coding scheme (Fig. \ref{codingschemechosen}) and modulation scheme (Fig. \ref{modulationschemechosen}) with the lowest transmission overhead (Fig. \ref{transoverhead}), the highest code rate (Fig. \ref{coderate}) and throughput (Fig. \ref{throughputTH}). Assume that the time when the VR user is idly standing  at any position is randomly chosen in  $[3,4,5,6,7]$ minutes. We assume $R$ redundant bits are generated from $K$ original bits so that the number of bits can be corrected to be $t_{MDPC}=1$ bit for MDPC code and to be $t_{RS}=1$ symbol for RS code. The reason for this assumption is because with a low redundant coding data, the coding and decoding time is optimized with a low latency, while it can still satisfy the optimized fault tolerance with high code rate and throughput, as discussed in section \ref{algo}. Assume that the processing time for the feedback controller and adaptive configuration module is small and ignored.

Fig. \ref{coderate}, Fig. \ref{transoverhead}, Fig. \ref{codingschemechosen}, Fig. \ref{modulationschemechosen} and Fig. \ref{throughputTH} theoretically show the code rate, transmission overhead, chosen coding and modulation scheme and throughput, respectively, where the throughput is observed in the time interval, when VR user idly stands at a specific position. $1$ and $2$ denote the chosen coding scheme, respectively. $4$, $5$, $6$ and $7$ denote the chosen modulation scheme of BPSK, QPSK, 8PSK and 16QAM, respectively. Based on the standing positions of VR user predicted in Fig. \ref{distance} (line \ref{dis} in Algorithm \ref{algorithm1}), we can approximately estimate four bit error probabilities $p_e$ from the database of BER values over different transmission distances and modulation schemes in Fig. \ref{BER_1ch_BW8_64GHz}, stored at the adaptive configuration module, if THz BS uses BPSK, QPSK, 8PSK or 16QAM (line \ref{pe} in Algorithm \ref{algorithm1}). With four bit error probabilities from four modulation schemes if used by THz BS (line \ref{pe} in Algorithm \ref{algorithm1}), $t_{MDPC}=1$ bit and $t_{RS}=1$ symbol, we will find four values of $K$ for MDPC code satisfying Eq. \eqref{parameters-mn} and four values of $K$ and $s$ for RS code satisfying Eq. \eqref{parameters-RKforRS}. Theoretically, if we can choose a reasonable value of input data $K$ (section \ref{algo}), then the residual bit error probability $P_{re}^V \approx 0$ after decoding. Therefore, the throughput is equivalent to the information bit rate, i.e., based in Eq. \eqref{pa-rf-th}, $TH=R_F \cdot D$. With $R$ redundant bits deduced from $t_{MDPC}=1$ bit and $t_{RS}=1$ symbol by Eq. \eqref{tv}, and eight values of $K$ are previouly found, all of them are applied from Eq. \eqref{pa-rf-th}. As a result, the adaptive configuration module will find the best coding scheme (Fig. \ref{codingschemechosen}) and modulation scheme (Fig. \ref{modulationschemechosen}) with the highest code rate (Fig. \ref{coderate}) and throughput (Fig. \eqref{throughputTH}). The transmission overhead in Fig. \eqref{transoverhead} is given by: $\theta =1-R_F$, where $R_F$ is code rate.

For code rate in Fig. \ref{coderate}, $0.25 \leq R_F \leq 1$, i.e., $0 \leq \theta \leq 0.75$ in Fig. \ref{transoverhead}; the longer the transmission distance, the lower the code rate For the chosen coding scheme in Fig. \ref{codingschemechosen}, RS code performs better than MDPC. For the chosen modulation scheme in Fig. \ref{modulationschemechosen}, 16QAM is the best, while BPSK is the lowest selection because its data transmission rate is the lowest. We see that no modulation scheme 8PSK is chosen in our scenario. This can be explained since BER of 8PSK and 16QAM are quite similar, while the data transmission rate of 16QAM is higher than that of 8PSK. Therefore Algorithm \ref{algorithm1} always chooses the coding and modulation scheme with the code rate and throughput as the highest. As discussed above, with the suitable code rates or transmission overheads chosen from the best coding and modulation schemes in Fig. \ref{coderate} and Fig. \ref{transoverhead} over different channel states, we can theoretically improve the system fault tolerance with the residual bit error probability $P_{re}^V\approx 0$. Also, since $P_{re}^V\approx 0$, the throughput can achieve the information bit rate $TH=R_F \cdot D$, and due to the flexibility in choosing the best coding and modulation schemes with the highest code rate, sometime we can theoretically get a maximum throughput of $28.16$ Gbps in Fig. \ref{throughputTH}, equivalent to data transmission rate of 16QAM. 
 \section{Conclusion}
We proposed a THz data and control plane system architecture for VR applications in smart factories. The results shows that an adaptive design system can improve throughput and fault tolerance, whereby the residual bit error probability is $P_{re}^V \approx 0$ after decoding and we get can theoretically get a maximum throughput of $28.16$ Gbps, equivalent to data transmission rate of 16QAM. Future work needs to address algorithm computation offloading and its response time. 
\section*{Acknowledgment}
The authors acknowledge the financial support by the Federal Ministry of Education and Research, Germany, program "Souveran. Digital. Vernetzt." project 6G-RIC, 16KISK031 and partial support by the DFG Project Nr. JU2757/12-1.
\bibliographystyle{IEEEtran}

\bibliography{nc-rest}

\begin{thebibliography}{10}
\providecommand{\url}[1]{#1}
\csname url@samestyle\endcsname
\providecommand{\newblock}{\relax}
\providecommand{\bibinfo}[2]{#2}
\providecommand{\BIBentrySTDinterwordspacing}{\spaceskip=0pt\relax}
\providecommand{\BIBentryALTinterwordstretchfactor}{4}
\providecommand{\BIBentryALTinterwordspacing}{\spaceskip=\fontdimen2\font plus
\BIBentryALTinterwordstretchfactor\fontdimen3\font minus
  \fontdimen4\font\relax}
\providecommand{\BIBforeignlanguage}[2]{{%
\expandafter\ifx\csname l@#1\endcsname\relax
\typeout{** WARNING: IEEEtran.bst: No hyphenation pattern has been}%
\typeout{** loaded for the language `#1'. Using the pattern for}%
\typeout{** the default language instead.}%
\else
\language=\csname l@#1\endcsname
\fi
#2}}
\providecommand{\BIBdecl}{\relax}
\BIBdecl

\bibitem{boeing2018}
\BIBentryALTinterwordspacing
Boeing, ``Boeing tests augmented reality in the factory.'' [Online]. Available:
  \url{https://www.boeing.com/features/2018/01/augmented-reality-01-18.page}
\BIBentrySTDinterwordspacing

\bibitem{7444891}
F.~{Moshir} and S.~{Singh}, ``Rate adaptation for terahertz communications,''
  in \emph{2016 13th IEEE Annual Consumer Communications Networking Conference
  (CCNC)}, Jan 2016, pp. 816--819.

\bibitem{2018_WEHN.SAHIN_NextGenerationChannelCoding}
N.~Wehn and O.~Sahin, ``Next-{{Generation Channel Coding Towards Terabit}}/{{S
  Wireless Communications}},'' Aug. 2018.

\bibitem{wehn_norbert_2018_1346686}
\BIBentryALTinterwordspacing
\emph{{Next-Generation Channel Coding Towards Terabit/s Wireless
  Communications}}.\hskip 1em plus 0.5em minus 0.4em\relax Zenodo, Aug. 2018.
  [Online]. Available: \url{https://doi.org/10.5281/zenodo.1346686}
\BIBentrySTDinterwordspacing

\bibitem{wehn_norbert_2019_3360520}
\BIBentryALTinterwordspacing
N.~Wehn, ``Channel coding for tb/s communications,'' Aug. 2019. [Online].
  Available: \url{https://doi.org/10.5281/zenodo.3360520}
\BIBentrySTDinterwordspacing

\bibitem{8292607}
S.~Nie and I.~F. Akyildiz, ``Three-dimensional dynamic channel modeling and
  tracking for terahertz band indoor communications,'' in \emph{2017 IEEE 28th
  Annual International Symposium on Personal, Indoor, and Mobile Radio
  Communications (PIMRC)}, 2017, pp. 1--5.

\bibitem{9013838}
R.~Singh and D.~Sicker, ``Parameter modeling for small-scale mobility in indoor
  thz communication,'' in \emph{2019 IEEE Global Communications Conference
  (GLOBECOM)}, 2019, pp. 1--6.

\bibitem{DBLP:journals/twc/PengGK18}
B.~Peng, K.~Guan, and T.~K{\"{u}}rner, ``Cooperative dynamic angle of arrival
  estimation considering space-time correlations for terahertz
  communications,'' \emph{{IEEE} Trans. Wirel. Commun.}, vol.~17, no.~9, pp.
  6029--6041, 2018.

\bibitem{Chamania2020}
\BIBentryALTinterwordspacing
M.~Chamania and A.~Jukan, \emph{Dynamic Control of Optical Networks}.\hskip 1em
  plus 0.5em minus 0.4em\relax Cham: Springer International Publishing, 2020,
  pp. 535--552. [Online]. Available:
  \url{https://doi.org/10.1007/978-3-030-16250-4_15}
\BIBentrySTDinterwordspacing

\bibitem{8867037}
I.~Vertat and L.~Dudacek, ``Multidimensional cross parity check codes as a
  promising solution to cubesat low data rate downlinks,'' in \emph{2019
  International Conference on Applied Electronics (AE)}, 2019, pp. 1--5.

\bibitem{Reed1999}
I.~S. Reed and X.~Chen, \emph{Reed-Solomon Codes}.\hskip 1em plus 0.5em minus
  0.4em\relax Boston, MA: Springer US, 1999, pp. 233--284.

\bibitem{9391508}
B.~K. Jung, C.~Herold, J.~M. Eckhardt, and T.~Kürner, ``Link-level and
  system-level simulation of 300 ghz wireless backhaul links,'' in \emph{2020
  International Symposium on Antennas and Propagation (ISAP)}, 2021, pp.
  619--620.

\end{thebibliography}

\end{document}